# PROJECT REPORT

**ON**

# SOME STUDIES ON THE EFFECT OF SIZE ON VARIOUS SEMICONDUCTOR STRUCTURE USED IN DEVICE DESIGN.

**PROJECT WORK PERFORMED BY**

**JATINDRANATH GAIN**
**ASSISTANT PROFESSORE**
**DEPARTMENT OF PHYSICS**
**DEROZIO MEMORIAL COLLEGE**

**PROJECT WORK SUPERVISED BY**

**PROF. SUDAKSHINA KUNDU**
**HOD- department of computer science & engineering**
**West Bengal University of Technology**
**BF-142, Salt Lake Sector-1, Calautta-700064**

**PROJECT WORK SPONSORED BY**

**UNIVERSITY GRANT COMMISSION**
**INDIA**

# CONTENTS:

# 1.Abstract :

The quantum mechanical tunneling through multiple quantum barriers is a long-standing and well-known problem. Three methods proposed earlier to calculate the tunneling probabilities and energy splitting: (1). Instanton Method (2) WKb Approximation (3) Numerical Calculation.Instaton method is helpful to understand the physical insight of quantum tunneling but the validity is restricted to the case of large separation between the two potential minima. WKB approximation is widely used in its simple mathematical form, but the result is inaccurate due to its inherent defect in connection formula. Recently WKB approximation has been developed by changing the phase lose at the classical turning points but no above approximation have provide the perfect result to the best of knowledge of Author. Using numerical methods, one can get the solution up to the desired accuracy, but a considerable deal of physical insight is lost in this process. In this paper, the Author presented the developmentof models of multiple quantum wells or barriers potential by using analytical Transfer matrix method (TMM), which has been applied to any arbitrary potential wells and barriers successfully. The author applied the above theory to three electronic device models and got satisfactory results.

# 2.Introduction:

Low dimensional carrier systems in the semiconductor heterostructures are gaining much importance in recent times due to the potential use of their unique properties in applications ranging from optoeltronics to high speed devices [1-4]. In this connection perpendicular transport of the carriers in semiconductor heterostructures has attracted much attention [5-8]. The MQW structures in particular, are becoming very important due to their potential use in the design and fabrication of quantum cascade lasers, resonant photo detectors, resonant tunneling diodes, single electron tunneling transistors [8] etc. Moreover, with the decrease in the dimensions of the CMOS devices the effect of tunneling of carriers becomes very important in order for estimating the various leakage currents flowing through the devices present in the VLSI chips.

In this paper an attempt has been made to study the tunneling of carriers through a quantum barrier and plot the variation of the transport coefficient with respect to carrier energy. The range of energy include the classically forbidden transitions as well. The transmission/tunnelling coefficient, which is the flux of particles penetrating through the potential barriers to the flux of particles incident on it at the other interface, has been computed by using the Ben Daniel-Duke (BDD) boundary conditions [9] for solving the Schrödinger equations for the electrons (carriers in this case) inside the coupled well regions and the barrier in between. The theory has been based on the transfer matrix method. Tunneling depends significantly on the barrier width. Scaling of structure dimension affects this variation very sharply.

The material pairs of interest include CdS/CdSe, AlGaAs/GaAs and InAs/AlSb. InAs/AlSbs based HEMTs are excellent for application in satellites due to their low operating voltages [3]. Devices based on AlGaAs/GaAs have been in use over quite some time and CdS/CdSe QW structures promise of improved gain performance for light emitting applications [2]

The effective masses of the carriers are different inside the well and in the barrier, which are made of different materials. Further, this effective mass may change with energy as is given in the case of the InAs/AlSb pair [10]. This affects the tunneling behaviour of the electrons because there is a dependence of the transmission coefficient of the carrier effective masses. In this paper the effect of variation in effective mass on transport coefficient has also been studied. The variation of transport coefficient with electron energy has been studied and compared for different barrier widths for each of these material pairs.

Low dimensional carrier systems in the semiconductor heterostructures are gaining much importance in recent times due to the potential use of their unique properties in applications ranging from optoeltronics to high speed devices [1-4]. In this conection perpendicular transport of the carriers in semiconductor heterostructures has attracted much attention [5-8]. The MQW structures in particular, are becoming very important due to their potential use in the design and fabrication of quantum cascade lasers, Quantum Infra-red photo-detectors (QWIP), Quantum cascade detectot (QCD), resonant photodetectors, resonant tunneling diodes, single electron tunneling transistors [8] etc.  Recently much research is carried on in the field of semiconductor infrared detectors. These are based on inter-subband transitions and have the possibility of device tailoring for distinct, narrow band detection energy range.  Quantum cascade detectors (QCD)s are based on a bound-to-bound transition. Here the excited state is coupled to an extraction cascade / phonon-stair that transport electrons vertically to the ground state of the next period that follows. [9]

In this paper an attempt has been made to derive a general mathematical model that will help in evaluating the energy Eigen values inside asymmetric and aperiodic MQW structures. The model is tested on the experimental data for a 16.5 μm QCD of reference 9. This model will also help in studying the tunneling of carriers one  well to another through a quantum barrier. The transmission coefficient, which is the flux of particles penetrating the potential barrier to the flux of particles incident on it at the other interface, has been computed by using the Ben Daniel-Duke (BDD) boundary condition [9] for solving the Schrödinger equations for the electrons (carriers in this case) inside the coupled well regions and the barrier in between. The theory has been based on the transfer matrix method [10-16]. Tunneling depends significantly on the barrier width. Scaling of structure dimension affects this variation very sharply.

The structure on which the mathematical model is tested is an array of $In_{.53}Gs_{.47}As/In_{.52}Al_{.48}As$ layers of thicknesses in Angstrom in the order 157/**55**/69/**26**/90/**43**/94/**40**/95/**38**/96/**28** [17]. The first two wells have widely different widths and the barriers that separate them and flank the pair on either side are of unequal width. The effective mass of the carriers are different inside the well and in the barrier, which are made of different materials. The model is tested on this well pair to match the computed result of coupled energy states with those obtained experimentally. The agreement is satisfactory.

Inside the structure the wells and barriers of widths 90/**43**/94/**40**/95/**38**/96 can be approximated by a periodic pattern as the wells widths are varying very slightly and so do the barrier width. Under this circumstance the general model can be simplified and the energies are computed. The comparison with experimental data prove encouraging.

## 3.(a) Methodology – 1st Phase:

The quantum mechanical theory of tunneling through a classically forbidden energy state can be extended for other types of classically forbidden transitions. In this paper a generalized theory of quantum tunneling for transition through multiple quantum barriers has been developed and tested for three different pairs of materials. These adjacent lower energy regions, that are separated by a quantum barrier, are coupled and this is the general pattern for many of the heterostructures. The electron wave functions in the lower energy regions and the barrier region are obtained by solving the Schrödinger equations that satisfy appropriate boundary conditions [11,12]. The solutions depend on the effective masses of the carriers in the regions concerned. Hence the tunneling probability will not only be a function of the dimensions of the barrier alone but will be affected by change of materials as well as its energy dependent effective mass.

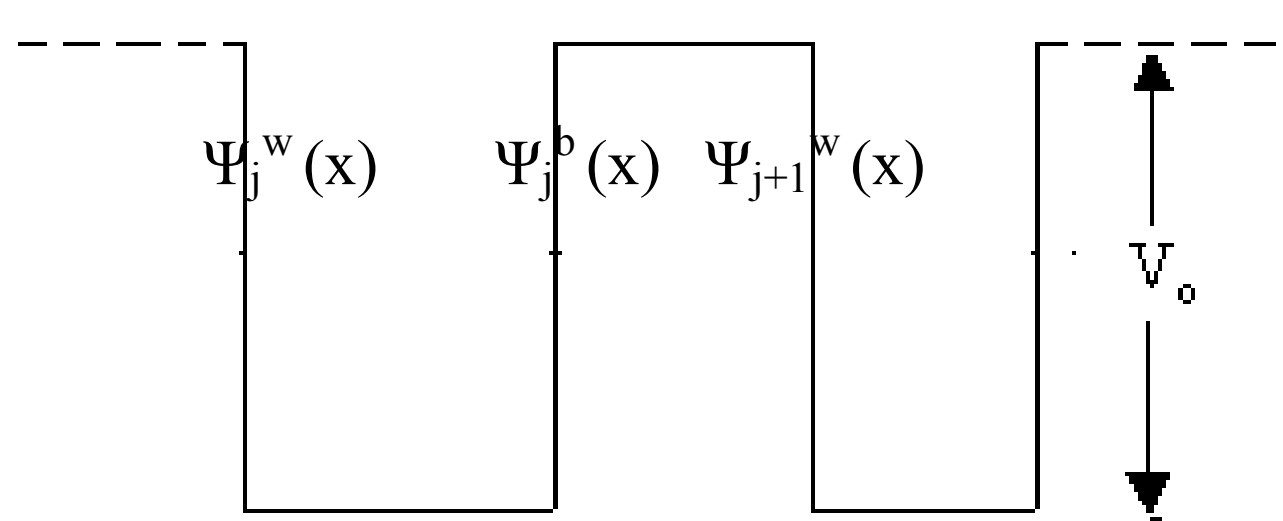

Fig.1.Schematic illustration of a multiple quantum wells (MQW) heterostructure.

The total wave function $\Phi(r,x)$ for the electron can be written as

$\Phi(r, x) = \exp(ik_0 r)\, \Psi(x)$ 1

Where $\Psi(x)$ satisfies the one dimensional Schrödinger equation with position-dependent electron effective Mass, x represents the growth direction of heterostructure.

The Schrödinger equation according to the effective mass theory for the finite potential barrier and the well regions on either side take on the well-known form

$$\frac{\hbar^2}{2m^*}\frac{d^2\Psi(x)}{dx^2} + (E - V)\Psi(x) = 0 \qquad 2$$

with appropriate effective masses m* and potential energies V for the region where the equation is defined. Inside the barrier region the effective mass $m^* = m_B$ and potential energy is Vo while those outside the barrier are respectively given by $m^* = m_w$ and zero.

Now, we use the transfer matrix for the jth junction and w should generalize the result for N-junctions. This matrix relates the coefficients of the wave function at one end of the junction to those of the other one, so that the wave function can be written as

$\Psi_j^{w}(x) = A_j^{w}\exp(ik_w x) + B_j^{w}\exp(-ik_w x)$ 3

for the jth well, and

$\Psi_j^{b}(x) = A_j^{b}\exp(ik_b x) + B_j^{b}\exp(-ik_b x)$ 4

for the jth barrier

Where $k_w = \left(\frac{2m_w E}{\hbar^2}\right)^{\frac{1}{2}}$ and $k_b = \left[\frac{2m_b(V_0 - E)}{\hbar^2}\right]^{\frac{1}{2}}$ with $m_w$ and $m_B$ are the electron effective masses in the regions outside the barrier and barrier region respectively.

By matching the continuity of the wave function $\Psi(x)$ and its appropriate normalised derivative $\frac{1}{m^*}\frac{d\psi(x)}{dx}$ at the boundaries and form $2\times 2$ transfer matrix equations for each interface and we derive a matrix formula that relates the coefficients $A_j$ and $B_j$ with $A_{j+1}$ and $B_{j+1}$.

$$\begin{bmatrix} A_j \\ B_j \end{bmatrix} = \begin{bmatrix} M_{11}^{[j]} & M_{12}^{[j]} \\ M_{21}^{[j]} & M_{22}^{[j]} \end{bmatrix} \begin{bmatrix} A_{j+1} \\ B_{j+1} \end{bmatrix} \qquad 5$$

By using equation (5) we obtained the coefficients of the wave function at the leftmost slab to those of the right most slabs

$$\begin{bmatrix} A^j \\ B^j \end{bmatrix} = \frac{1}{2}\begin{bmatrix} 1 & -ik_w^{-1} \\ 1 & ik_w^{-1} \end{bmatrix} M_j \begin{bmatrix} 1 & 1 \\ ik_w & -ik_w \end{bmatrix} \begin{bmatrix} A_{j+1} \\ B_{j+1} \end{bmatrix} \qquad 6$$

Where $M_j$ is the jth transfer matrix corresponding to the jth junction written as:

$$M_j = M_b(b_j)M_w(a_j)M_b(b_{j+1}) \qquad 7$$

Where $b_j$ and $a_j$ are the widths of the jth barrier and jth well respectively, $M_b(b_j)$ and $M_w(a_j)$ Correspond to the transfer matrices for the jth barrier and jth well respectively.

Here the total transfer matrix is expressed as the cascading of a series of individual barrier and well.

From equation (6) & (7) it is readily found the transmission amplitude Q is given by

$$Q = \frac{2}{M_{11} + M_{22} + i(k_w M_{12} - k_w^{-1} M_{21})} \qquad 8$$

Where $M_{ij}$ are the elements of the total transfer matrix.

From the definition of transmission coefficient (T), we obtained the following expression from equation (8)

$$T = |Q|^2 \qquad 9$$

From equation (5) we obtained the expansion coefficient of the wave function at the left most slab to those of the right most slab as:

$$\begin{bmatrix} A_1 \\ 0 \end{bmatrix} = M^{[1]}M^{[2]}M^{[3]}......M^{[j-1]}\begin{bmatrix} A_j \\ 0 \end{bmatrix} \qquad 10$$

$$= M_{total}\begin{bmatrix} A_j \\ 0 \end{bmatrix} \qquad 11$$

Which satisfies $M_{21} = 0$ 12

We obtained the energy Eigen values and Eigen functions by solving equation (12).

The wave vector $k_w$ outside the barrier is a real quantity for all positive energies E of the electron. The conduction band minimum in the region outside the barrier is taken as the zero of energy. Since the conduction band minimum of the barrier is above that of the region outside hence for certain energies of the electron (E<Vo) the wave vector $k_b$ will be imaginary and for energies above Vo (E>Vo) it will be real. Thus the energy is divided into two regions, (E<Vo) for non-classical transition by tunneling and (E>Vo) where transition can take place even under classical conditions. The solutions will be different and the methods applied to evaluate the transmission coefficients in the two regions will be different. Since $k_w$ and $k_b$ are both dependent on effective mass and energy, hence for different material pairs the variation of the transmission coefficient for energy values normalized with respect to the barrier height will be different.[13,14]

Here $\beta = m_B / m_w$ is called the mass discontinuity factor. It plays a very important role in determining the transmission coefficient.

For perfect transmission through a barrier sandwiched between two wells we must have

$$\beta (k_b/k_w) - k_w/(\beta k_b)]^2 \sinh^2 k_w L = 0 \qquad 13$$

where L is the length of the barrier. So there will be transmission at energy values $E_n = Vo + (n\pi\hbar/L)^2/2m_B$ where transmission coefficient becomes 1. The barrier becomes transparent at these energy values. This is called Ramsauer-Townse [13] effect in atomic physics. The resonance condition is satisfied at normalized energy values $(E/Vo) = 1+n^2/\beta$, where n is an integer. The resonance values therefore depend upon the mass discontinuity factor.

When the electron energy E is less than the barrier height Vo the situation is somewhat different. The above derivations assume that the mass discontinuity factor remains unchanged with energy.

However, it has been pointed out [15] that for InAs/AlSb material system the effective mass varies with the energy according to the expression

$$m(E) = m^*[1+(E-E_c)/E_{eff}] \qquad 14$$

With $E_c$ representing the conduction band minimum and $E_{eff}$ is the effective bandgap at the energy E. The transmission coefficient normally has a dependence on energy through the expression of the wave vector $k_b$. The energy variation of the mass discontinuity factor $\beta(E)$ adds another dimension to this variation. It is worthwhile to examine the effect of this variation and how far it affects the tunneling phenomenon.

The transmission coefficient of electrons through a potential barriers is important for studying the leakage current in MOSFETs with dimensions in the nanometer range. It is also a crucial parameter for studying the behavior of multiple quantum well structures where the barrier is sandwiched between two coupled quantum wells. In the second case the equations are modified to include the well dimensions also. When both the well and barrier regions are in the nanometer range we expect further quantization of the energy levels. This is being considered in a further study of the multiple quantum well structure.

# 3. (b) Methodology – 2$^{nd}$ Phase:

The quantum mechanical theory of tunneling through a classically forbidden energy state can be extended for other types of classically forbidden transitions. In this paper a generalised theory of quantum tunneling for transition through a barrier between two quantum wells has been developed where the electron wave functions in the well region and the barrier region are obtained by solving the Schrödinger equations that satisfy appropriate boundary conditions [10,11]. The solutions depend on the effective masses of the carriers in the regions concerned and the height of the barrier on either side of the well. Hence the tunneling probability will not only be a function of the dimensions of the barrier alone but will be affected by change of materials as well.

We have started with the most generalised case where the well widths and barrier widths are all unequal. In addition to this the well and barrier materials can be different with different barrier hights and effective electron masses. If we assume that the barrier widths are such that adjacent wells are coupled through the barriers then we will have to deal with five wave functions; the wave functions of the electrons inside the wells and the electron wave functions for the barriers on either side of these wells. We consider the Schrodinger wave equations for a series of three barriers of widths 2d, 2b and 2f with wells of width 2a and 2c between the barriers 2d and 2b and 2b and 2f respectively. The well barrier combination is shown in figure 1 below.

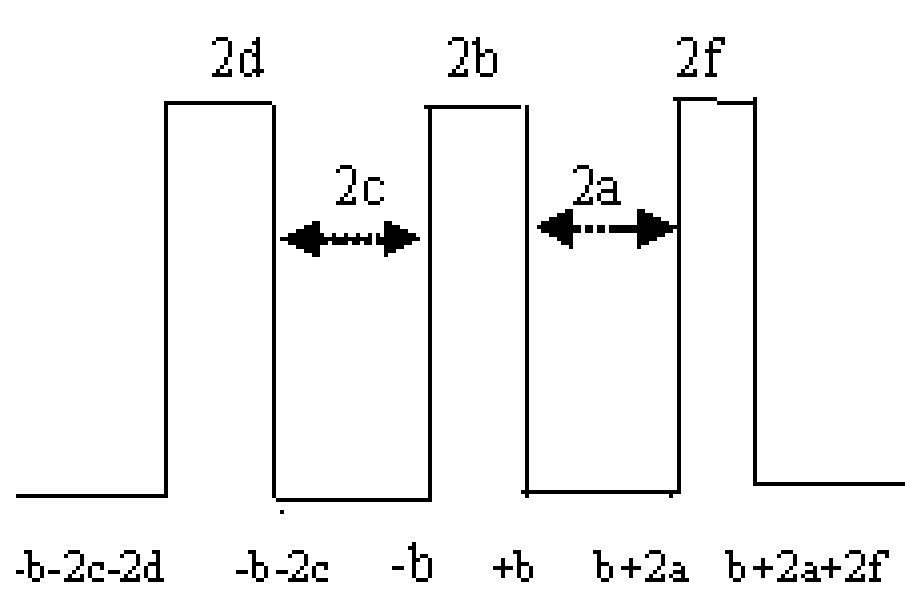

Figure 1: Araay of MQWs.

The Schrödinger equation according to the effective mass theory for the finite potential barrier and the well regions on either side take on the well known form

$$(\hbar^2/2m^*)\ d/dx\ \{d\Psi/dx\} + (E-V)\Psi = 0 \qquad 1.$$

with appropriate effective masses m* and potential energies V for the region where the equation is defined. Inside the barrier region the effective mass $m^* = m_B$ and potential energy is $V_o$ while those outside the barrier are respectively given by $m^* = m_w$ and zero.

Solution of the Schrödinger equation yields the electron wave functions inside the barrier and outside which is represented as for the most general case:

$$\Psi_{B1}(x) = A_{B1} \exp(ik_{B1}x) + B_{B1} \exp(-ik_{B1}x) \quad \text{for; } -(b+2c-2d)<x< -(b+2c) \quad \text{in the first barrier} \qquad 2.$$

$\Psi_{W1}(x) = A_{W1} \exp(ikw_1x) + B_{W1}\exp(-ik_{W1}x)$ for $-(b+2c) < x < -b$ 3.
Inside the first well

$\Psi_{B2}(x) = A_{B2} \exp(ik_{B2}x) + \mathbf{B_2} \exp(-ik_{B2}x)$ for $-b < x < +b$
in the second barrier 4.

$\Psi_{W1}(x) = A_{W2} \exp(ikw_2x) + D_2 \exp(-ik_{W2}x)$ for $b < x < b+2a$
inside the second well 5.

$\Psi_{B3}x) = A_{B3} \exp(ik_{B3}x) + B_{B3} \exp(-ik_{B3}x)$ for $(b+2a) < x < (b+2a+2f)$
in the second barrier 6.

Where $k_{Bn} = (2m_{Bn}[Vo_n - E]/\hbar^2)^{1/2}$ and $k_{Wi} = (2m_{Wi}E/\hbar^2)^{1/2}$ with $m_{wi}$ and $m_{Bn}$ are the electron effective masses in the well and barrier regions respectively. In the most general case the barrier heights $Vo_n$ and electron effective masses $m_{wi}$ and $m_{Bn}$ in the different regions are assumed to be all different.
The boundary conditions that are applied assume continuity of the wave functions at the well-barrier boundaries i.e., $\Psi_{Bn}(x) = \Psi_{Wi}(x)$ at each boundary.
The other important boundary condition called Ben Daniel Duke (BDD) boundary condition relates the flux of electrons incident on the barrier to that which penetrates it. This is expressed mathematically as

$(1/m_{wi})[d\Psi_{Wi}(x)/dx] = (1/m_{Bn})[d\Psi_{Bn}(x)/dx]$
at well-barrier interface. 7.

By matching the continuity of the wave function $\Psi(x)$ and its appropriate normalized derivative as described above the transfer matrix method can be applied to find out the transmission coefficient $\tau = I(F/A)^2 I$ at the farthest region. The Matrix equation is very complex and involves a large number of terms. The coupling energy between two adjacent wells can be obtained by putting $\tau = 1$ and varying the energy value to obtain the best fit. This iterative method will yield the energy for which electrons may be coupled to both the wells by penetrating through the intervening barrier.

This mathematical model is tested on an array of $In_{.53}Gs_{.47}As/In_{.52}Al_{.48}As$ layers of thicknesses in Angstrom in the order 157/**55**/69/**26**/90/**43**/94/**40**/95/**38**/96/**28** [17]. The first two wells have widely different widths and the barriers that separate them and flank the pair on either side are
9
of unequal width. The effective mass of the carriers are the same in both the wells ($m_{w1}=m_{w2}=m_w$) but different from that in the barriers, which are made of the same material so that $m_{B1}=m_{B2}=m_{B3}=m_B$. The barrier hights in all the barrier regions are same and equal to $V_o$. The transmission coefficient derived by Transfer Matrix method for the most general case simplifies to some extent and becomes

$$\tau = [4 k_B k_W / (m_w m_B)]^4 / [T_1 + 2T_2 + 2 T_3 + 2 T_4] \quad 8.$$

where

$$T_1 = \{(k_B / m_B) + (k_W / m_W)\}^8 + \{(k_B / m_B) - (k_W / m_W)\}^8 + 6\{(k_B / m_B)^2 - (k_W / m_W)^2\}^4 \quad 9.$$

$T_2 = \{(k_B / m_B) + (k_W / m_W)\}^4 \{(k_B / m_B)^2 - (k_W / m_W)^2\}^2 [\cos(4\{bk_B + ck_W\}) + \cos(4\{ak_B + bk_W\} - 2\cos(4ck_W) - \cos(4ak_B) - \cos(4bk_B) - \cos(4\{ak_W + bk_B + ck_W\})]$ 10.

$T_3 = \{(k_B / m_B)^2 - (k_W / m_W)^2\}^4 [4\cos(4ak_W)\cos(4ck_W) - 2\cos(4bk_B)\{1+\cos(4[a-c]k_W)\} - 2\{\cos(4ak_W) + \cos(4ck_W)\} 1 - \cos(4bk_B)\}$ 11.

$T_4 = \{(k_B / m_B) - (k_W / m_W)\}^4 \{(k_B / m_B)^2 - (k_W / m_W)^2\}^2 [\cos(4\{bk_B - ak_W\}) + \cos(4\{bk_B - ck_W\}) - \cos(4ak_W) - \cos(4bk_B) - \cos(4\{bk_B - ak_W - ck_W\})]$ 12.

In order to compute the coupling energies the transmission coefficient is equated to 1 and energy values are obtained by iterative method.

When the well widths and barrier widths are respectively same and the same material is used for the well regions and the barrier regions respectively then the equation for the transmission coefficient simplifies to equation 13.

$(m_W k_B)/(m_B k_W) = \tan(ck_B)/\tan(ck_W)$ where c is the well width. 13.

The energy values are computed by iterative method. The transmission coefficient takes the form given below.

$\tau = 16(k_B k_W)^2 / [(m_B m_W)^2 \{(k_B / m_B) + (k_W / m_W)\}^4 + \{(k_W / m_W) - (k_B / m_B)\}^4 - 2\{(k_W / m_W)^2 - (k_B / m_B)^2\}^2 \cos(4ck_B)\}]$ 14.

At the coupling energies obtained from equation 13 the transmission coefficients given by equation 14 will tend to unity.

# 4. (a) Results: 1st Phase:

In this section we present our results obtained numerically by using MATLAB programming for the transmission coefficient across quantum wire containing multi-barrier heterostructure. The pairs of materials chosen are CdS/CdSe, AlGaAs/GaAs and InAs/AlSb. The parameters used in the computation are given in table 1. The effective mass of electrons in $Al_xGa_{1-x}As$ depends on the mole fraction of x, where x represents the concentration of Al.

In case of $Al_xGa_{1-x}As$ the effective mass is given by $m_{AlGaAs} = (0.063+0.083x)$ and the energy band gap is $E_{AlGaAs} = (1.9+0.125x+0.143x^2)$.

For each pair three values of the barrier width L are taken; 5nm, 10nm and 20nm. In all cases the transmission coefficient increases with diminishing dimension, as expected. This increase is the slowest for InAs/AlSb, which has the highest value of β. The effective mass variation for AlSb/InAs pair is included in the calculations. It is found that the value of the mass discontinuity factor does not change very sharply with energy for AlSb/InAs. However this slow variation of mass discontinuity factor β appears to slow down the increase of transmission coefficient, especially at very low barrier width (5nm).

Table-1:

| Parameter | CdS/CdSe | $Al_xGa_{1-x}As/GaAs$ (x= 0.47) | AlSb/InAs |
|---|---|---|---|
| Conduction Band-gap $\Delta E_c$ | $Eg_{CdS}$: 2.36eV; $Eg_{CdSe}$: 1.69eV; $\Delta E_c$= 67% of $(Eg_{CdS} - Eg_{CdSe})$ = 0.45 eV | $Eg_{AlGaAs}$: 1.99eV; $Eg_{GaAs}$: 1.42eV $\Delta E_c$=67% x $(Eg_{AlGaAs} - Eg_{GaAs})$ = 0.38 eV | $Eco_{InAs}$: 0.0 eV; $Eco_{AlSb}$: 2.1eV; $\Delta E_c = (Eco_{AlSb} - Eco_{InAs})$ = 2.1 eV |
| Effective mass (m*) | $m^*_{CdS}$: 0.20$m_o$; $m^*_{CdSe}$: 0.13$m_o$ | $m^*_{GaAs}$: 0.067$m_o$; $m^*_{AlGaAs}$:0.106$m_o$ | $m^*_{InAS}$: 0.020$m_o$; $m^*_{AlSb}$: 0.098$m_o$ |
| Mass discontinuity | β = 1.54 | β =1.58 | β = 4.9 |

Quantum tunneling for all barrier widths is least for AlSb barrier in InAs/AlSb pair and most significant for AlGaAs barrier for the in GaAs/AlGaAs/GaAs. This appears to be quite justified because InAs/AlSb/InAs structure has the highest barrier height and GaAs/AlGaAs/GaAs the lowest.(Figures 2, 3, 4)

As the barrier width decreases the tunneling increases and transmission coefficient value rises with normalised electron energy. The rate of rise is sharpest for GaAs/AlGaAs/GaAs and rather slow for InAs/AlSb/InAs. This is easily explained if the barrier heights are compared. CdS/CdSe/CdS structure lies midway.

For ($E_{nor}$ = E/Vo) <1, the transmission coefficient increases from 0 to 1 in a non-linear fashion. For each pair, the transmission coefficient is lower for wider wells as expected. However, the rise is the slowest for InAs/AlSb/InAs because of the highest value of the mass discontinuity factor.

Beyond the normalized energy ($E_{nor}$ = E/Vo) >1, there is resonance; i.e., there are quantised energy values where transmission reaches peak values sharply. This variation is prominent for all three structures, especially near $E_{nor}$ = 1. As the barrier width decreases the peaks get more separated in energy; this is most prominent in GaAs/AlGaAs/GaAs structure. For AlSb/InAs/AlSb the difference between the maxima and minima of the values of the transmission coefficients remain relatively unchanged with decreasing well width. For CdSe/CdS/CdSe and GaAs/AlGaAs/GaAs the maxima of transmission coefficient gradually increase as $E_{nor}$ increases beyond 1.The difference between the maxima and minima also gradually decreases.

Here the regions of lower band gap are taken to be semi-infinite. Hence the normalized energy values vary continuously. If the width of the lower band gap materials outside the barrier are reduced to the order of nanometers then the energy values inside the quantum wells will be quantised. This effect will be reflected in the nature of variation of the transmission coefficients with normalised energy and is expected to change significantly. This will have a crucial effect on carrier tunneling in multiple quantum well structures and leakage current in field effect devices. Studies of this effect are being explored by the author.

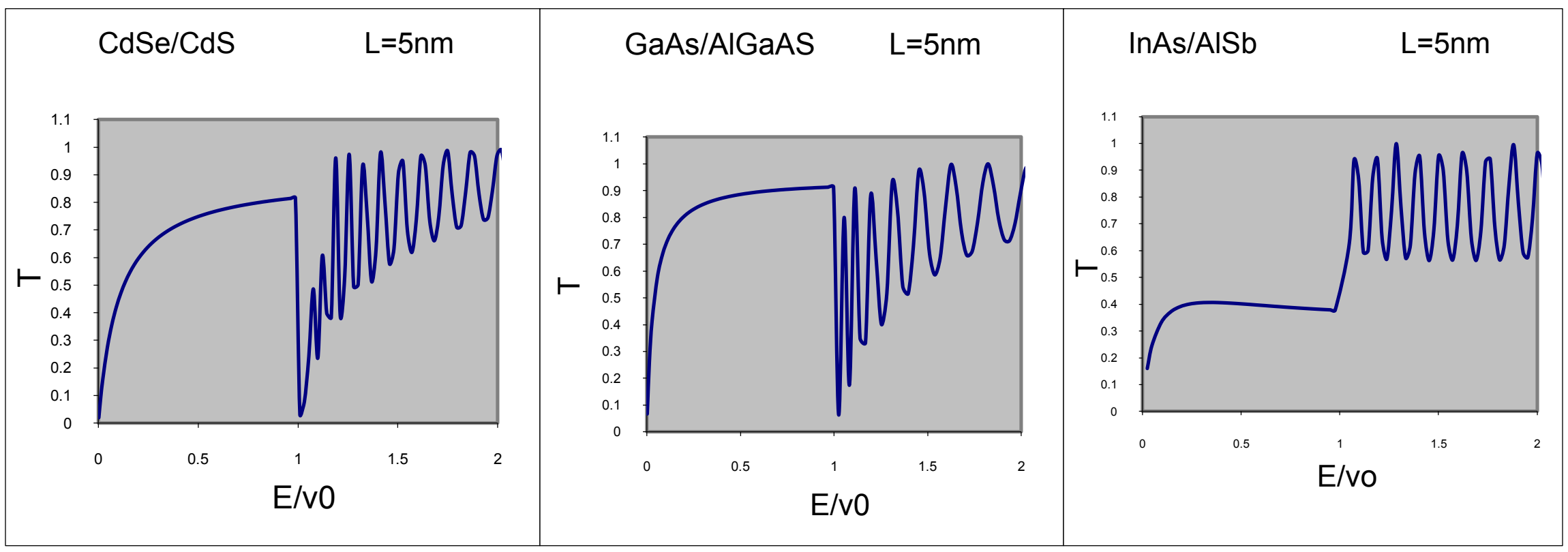

**(a)** **(b)** **(c)**

Fig.2: Variation of transmission coefficient of electrons with normalized energy E/Vo for (a) CdSe/Cds (b) GaAs/AlGaAs and (c) InAs/AlSb for well with.

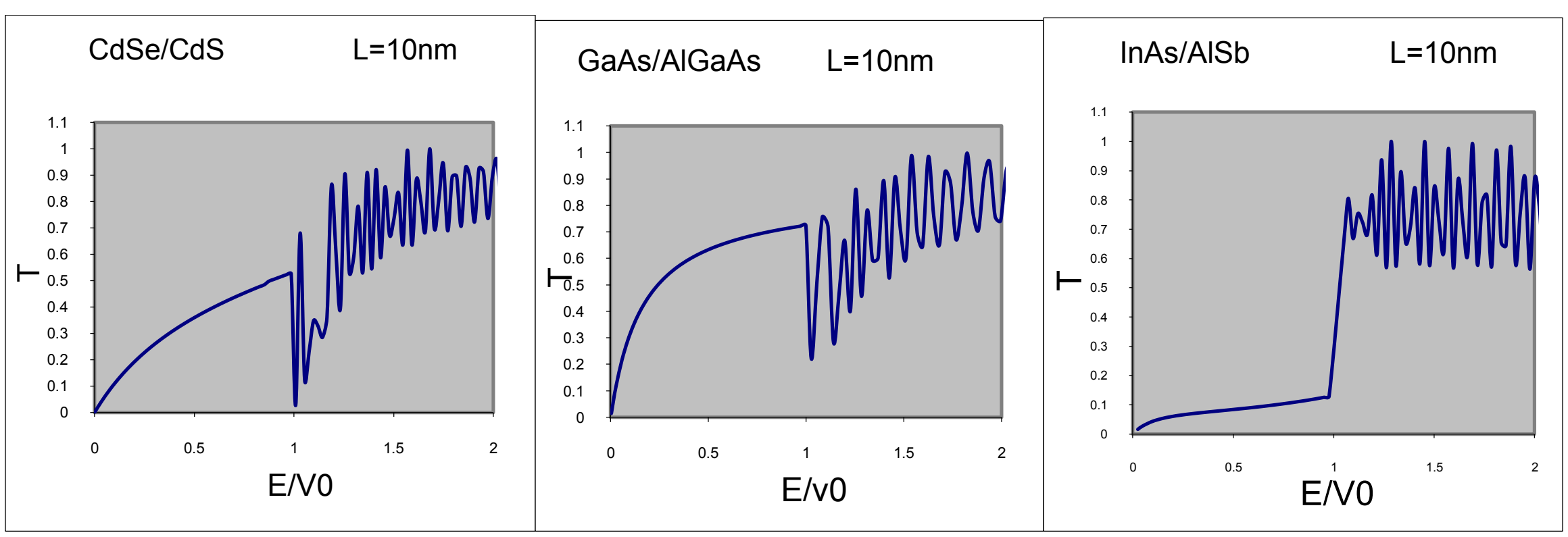

**(a)** **(b)** **(c)**

Fig.3: Variation of transmission coefficient of electrons with normalized energy E/Vo for (a) CdSe/Cds (b) GaAs/AlGaAs (c) InAs/AlSb for well with 10 nm.

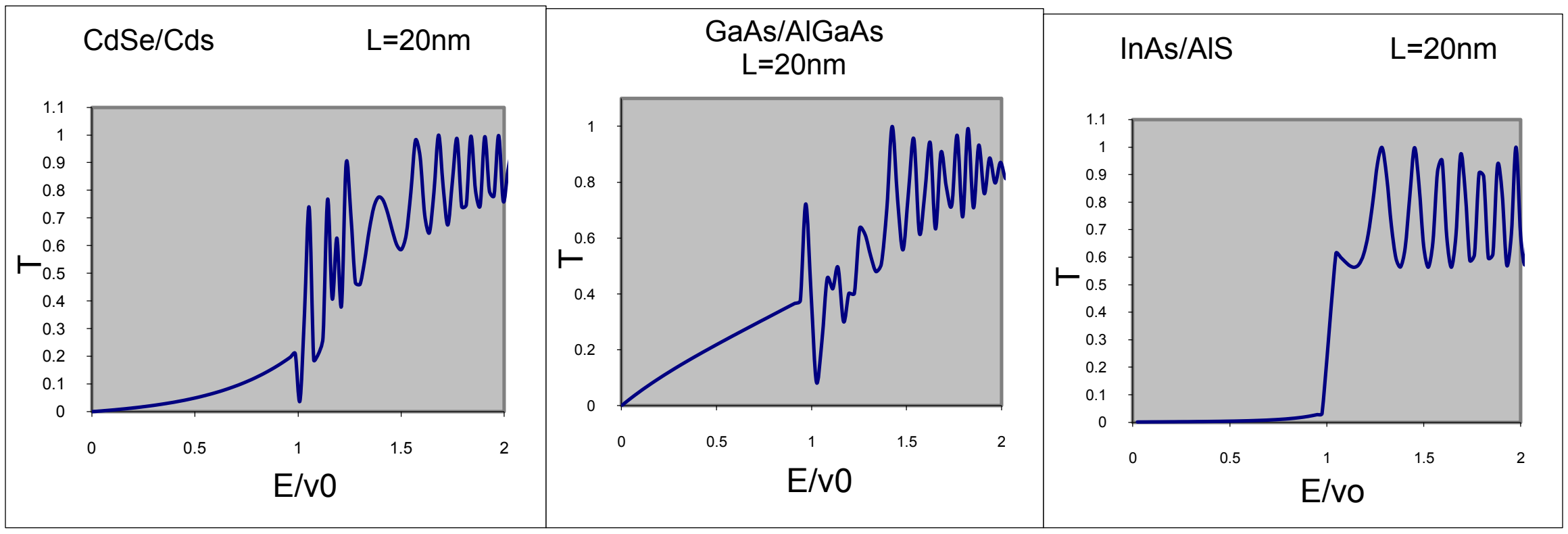

**(a)** **(b)** **( c)**

Fig.4: Variation of transmission coefficient of electrons with normalized energy E/Vo for (a) CdSe/Cds (b) GaAs/AlGaAs (c) InAs/AlSb for well with 20 nm

# 4. (b)Results: 2nd Phase:

The model is tested on a MQW structure obtained from a paper by F.R. Giorgetta et al [17]. The effective electron mass of the well material $In_{.53}Gs_{.47}As$ is taken to be $m_W = 0.043m_o$ [18] and that of the barrier layer of $In_{.52}Al_{.48}As$ is $m_B = 0.072m_o$. The well and barrier widths that alternate are given in Angstroms as 157/**55**/69/**26**/90/**43**/94/**40**/95/**38**/96/**28** [17]. The conduction band offset for $In_{.53}Gs_{.47}As/In_{.52}Al_{.48}As$ pair is 0.51eV.[17]

The array is divided in to two parts for testing the model. The first part consists of the first pair of wells of different widths lying between the three unequal barriers. The energies for which the transmission probabilities are unity are $E_1$=0.0205 eV, $E_2$=0.1014eV, $E_3$=0.1959eV, $E_4$=0.2899eV and $E_5$=0.3910eV. These computed values agree quite well with the experimental values given in figure 1 of reference 17. The experimental values are around 0.02eV, 0.10eV, 0.20 eV, 0.30 eV, 0.40eV and 0.45eV. The values cannot be determined with greater accuracy as they are not quoted but shown as the possible Eigen values corresponding to the Eigen functions. Tunnelling at around 0.2eV is very distinctly indicated. In our computation the transmission coefficient is 0.99 at 0.205 eV. For the other energies transmission coefficients are high but less than what we have for 0.205eV. The transmission coefficients at these energies are around 0.97.

The model is applied to the last three wells. Here the array of $In_{.53}Gs_{.47}As/In_{.52}Al_{.48}As$ pair is 90/**43**/94/**40**/95/**38** Angstroms. The well widths of 90**,** 94 and 95 Angstroms are quite close. So are the barrier widths of 43, 40 and 38 Angstroms. Here the simpler expressions represented by equations 13 and 14 may be applied without much error. The energies

Computed are 0.78 eV and 0.205 eV. The transmission coefficients for tunneling approach 0.99.

# 5. Conclusions:

In conclusion we may add that the general model developed can be used for designing MQW structures. To achieve tunneling through the structure at any specific energy value the well widths can be computed iteratively for $\tau = 1$at the specified energy. Applying this model to design MQW structures will be done in future research work.

**Acknowledgement:** -This work was financially supported by the University Grant Commission (UGC), India

# 6. References:

1. Fabrizio R.Giorgetta, Petter Krotz and Guido Sonnabend,"16.5nm quantum cascade detector using miniband transport" ,Applied Physics Letters ,Vol 90, 231111, **June-2007**

2. A.Hamed Majedi," Multilayer Josephson junction as a multiple quantum well Structure", IEEE transactions on applied superconductivity, Vol-17, No: 2, June

3. Manish K. Bafna, P. Sen. and P. K. Sen. (2006) "Effect of temperature dependence on nonlinear optical properties of InGaAs/GaAs single quantum dot" Indian Journal of Pure and Applied Physics, Vol. 42, pp. 949.**2007.**

4. C.Koeniguer, A.Gomez, V.Berger," Quantum cascade detector at 5 micrometers" ITQW-**2007**

5. Jeong S.Moon, Rajesh Rajavel, Steven Bui, Danny Wong and David H.Chow, "Room-temperature in InAlAs/InGaAs/InP Planer resonant tunneling couple Transistor" Applied Physics Letters 87. 183110 **(2005**).

6. D. Barate, R.Teissier, Y.Wang and A.N.Baranov, "Short wavelength intersubband emission from INAS/AlSb quantum cascade Structures", JAP-July-**2005**.

7. C.S.Ho, J.J.Liou, K.Y, Hung &C.C.Cheng."An analytical sub threshold current model for pocket implanted n MOSFET's", IEEE Trans. Electron Devices, VOl.50, Pp.1475-1479, June-**2003**.

8. Ryvkin BS, Avrutin EA, Pessa M. " Spontaneous emission, light-current characteristics, and polarization bistability range in vertical-cavity surface-emitting lasers. "J. Appl. Phys., Vol. 94, No. 7, pp 4267-4272, Oct, **2003.**

9. Agis A., Iliad's and A.Christou, Invited paper: 2003, Photonics West, Jan29-Feb03, **2003**.

8. E.Vittoz and J.Fellrath,"CMOS analog integrated Circuits based on Weak inversion Operation", IEEE journal of solid state circuits, Vol. Sc.-12, pp.224-231, June-1977.

9. V.T. Dolgopolov, A.A.Shashkin, and E.V.Deviatov,"Electron subbands in a double Quantum well in a quantizing magnetic field", Physical review B –15th May –1999

10. Semiconductor Optoelectronics Devices: Ed. P. Bhattacharya, Pub; Upper Saddle River; Prentice Hall, 2nd Edition, 1997.

11. Computational Microelectronics, Ed. S.Selberherr, Springer Wien New York, 2004.

12. Electron Transport in compound Semiconductor; B.R.Nag Springer-Verlag, Berlin Heidelberg Newyourk 1980

13. Nanotechnology and Nanoelctronics Material Devices Measurement Technique-Ed. W.R. Fahrner, Springer, 2004

14. R.Venugopal, Z.Ren, S.Dutta, M.S.Lundstron &D.Javanovic, J.Applied physics, 92, Pp-3730-3739 (2002).

15. O.Mann, C.L.Aravinda, and W.Freyland, Microscopic and "Electronic structure of Semimetallic Sb and Semiconducting AlSb Fabricated by Nanoscale Electrodeposition: An in Situ Scanning probe Investigation"- American chemical Society-2006.

16. Feng Zhou et. Al "Energy splitting double-well potential" Physical Review A 67. 062112 (2003).

17. Celine Joseph et.al "Scheme of Quantum Well Infrared Photo detector (QWIP) for Astronomical Application" International Conference on Sensors and Related Networks (SENNET'07) pp 153-157.

18. Jatindranath Gain et.al "General Model for the Computation of Energy Eigen Values in Aperiodic Multiple Quantum Well Structure" AOMD-2008, pp 102-107.

19. Theory of Semiconductor Devices, K.Hess, IEEE Press, 2000

20. Y.Tsividis, Operation and modeling of the MOS transistors, 2nd Edition, Singapore, McGraw-Hill, 1999.

21. Physics of Quantum Devices, B.R.Nag,-Klewer Academic Publishers.

22. Quantum Heterostructure, microelctronics and optoelectronics, -Vladimir, V.Matin.V.A.Kochelap, M.A.Stroscio,-Cambridge University Press.

23. Theory of Optical Processes of in Semiconductors, Bulk and Microstructures,-P.K.Basu, Oxford science Publication, 1997.

24. The Physics of Semiconductors with Applications to optoelectronic Devices- Kevin F. Brennan, Cambridge University Press

25. Quantum Mechanical Tunneling and Its Applications, D.K.Roy, World Scientific.

26. Optoelectronic Properties of Semiconductors and Super lattices-M.O. Manasreh, Series editor "Semiconductor Quantum wells Intermixing"-Vol-8, Edited by E. Herbert Li.

## 7.ACKNOWLEDGEMENT

*I would like to thank the Minor Research Project (MRP) program of University Grant Commission (UGC), India for its generous financial support*.

My sincere thanks to Dr. Biplob Chakroborty, Principal Derozio Memorial College to encourage me to doing this project & my PhD work.

I am very grateful to Prof. Sudakshina Kundu, HOD- Department of Computer Science & Engineering, West Bengal University of Technology for her constant guidance, encouragement & advice throughout the project work & cooperation.

I would like to express my hearty gratitude to Prof. Ghosh, Department of mathematics WBUT, Dr. Modhumita Das Sarkar, Department of Computer Science & Engineering, WBUT & Prof. Goutam Choudhury, Department of mathematics, Derozio Memorial College for their constant help & encouragement.

Last but not least my deepest regards to my father & mother & well-wishers who have given me inspirations for the hurdles of life

JATINDRANTH GAIN

Assistant Professore

Department Of Physics

Derozio Memorial College